\documentclass[twocolumn]{aastex7}
\received{March 11, 2025}

\submitjournal{ApJ}

\graphicspath{{./}{figs/}}

\usepackage{amsmath}
\usepackage{gensymb}

\shorttitle{Extreme Jet Beaming Observed in Neutrino-Associated Blazars}
\shortauthors{Plavin, Kovalev, \& Troitsky}

\begin{document}

\defcitealias{Plavin20}{P20}
\defcitealias{Plavin23}{P23}

\title{
Extreme Jet Beaming Observed in Neutrino-Associated Blazars
}

\correspondingauthor{A.~V.~Plavin}
\email{alexander@plav.in}

\author[0000-0003-2914-8554]{A.~V.~Plavin}
\email{alexander@plav.in}
\affiliation{Black Hole Initiative, Harvard University, 20 Garden St, Cambridge, MA 02138, USA}

\author[0000-0001-9303-3263]{Y.~Y.~Kovalev}
\email{yykovalev@gmail.com}
\affiliation{Max-Planck-Institut f\"ur Radioastronomie, Auf dem H\"ugel 69, Bonn D-53121, Germany}

\author[0000-0001-6917-6600]{S.~V.~Troitsky}
\email{sergey.troitsky@gmail.com}
\affiliation{Institute for Nuclear Research of the Russian Academy of Sciences, 60th October Anniversary Prospect 7a, Moscow 117312, Russia}
\affiliation{Physics Department, Lomonosov Moscow State University, 1-2 Leninskie Gory, Moscow 119991, Russia}

\begin{abstract}
Bright blazars were found to be prominent neutrino sources, and a number of IceCube events were associated with them over recent years. A particularly strong observational connection is present between neutrinos and blazars with bright, Doppler-boosted, parsec-scale radio emission. In this work, we further explore the nature of this connection by examining the jet geometry and kinematics of neutrino-associated blazars. We find that these blazars demonstrate remarkably strong jet beaming, even compared to other radio-bright sources. Their Doppler and Lorentz factors are larger, and viewing angles are smaller than for other blazars in the complete uniformly selected MOJAVE sample. Observationally, this serves as yet another piece of evidence for blazars forming a major population of neutrino sources. The strong neutrino-beaming correlation indicates that high-energy neutrino velocity is predominantly oriented along the jet, and the original PeV-scale protons exhibit a relativistic bulk motion along the jet. It suggests that neutrino production happens not too close to the black hole, but rather at sub-parsec distances, where the jet is already accelerated.
\end{abstract}


\section{Introduction}

Blazars were the first high-energy (TeVs-PeVs) astrophysical neutrino sources to be discovered \citep{IceCube-0506HE}. Later, neutrinos have also been detected from other sources, for instance the Milky Way Galaxy \citep{2022ApJ...940L..41K,2023Sci...380.1338I,2023PhLB..84137951A,2024arXiv241105608A} and nearby Seyfert-type Active Galactic Nuclei \citep[AGNs,][]{IceCubeNGC1068,NeronovSeyferts}. Still, blazars remain a very prominent class of bright neutrino emitters. There are more and more population-based studies resulting in neutrinos associated with AGNs selected by their compact radio flux (basically, blazars; e.g., \citealt{Plavin23,OVRO_IC_24,2024ApJ...964....3A,SurayNu}), X-ray flux (\citealt{2024PhRvD.110l3014K,NeronovSeyferts}; blazars in \citealt{Plavin24}), or using more specialized blazar selection methods \citep[e.g.,][]{2016NatPh..12..807K,2020MNRAS.497..865G,2022ApJ...933L..43B}. In this work, we adopt the commonly used definition of ``blazar'', referring to an AGN with a powerful jet aligned close to our line of sight.

Recent studies (e.g.~\citealt{Plavin23,OVRO_IC_24}) have specifically demonstrated a strong observational connection of radio-bright blazars being also bright neutrino sources. This connection is manifested both as the neutrino --~pc-scale radio flux correlation (see \autoref{s:significance} for updates), and also as a temporal correlation with neutrinos coming predominantly during radio flares. Those results already provide an indication that neutrinos are emitted predominantly along the jet axis in blazars; more direct studies should help quantifying this effect.
Furthermore, note that the strong neutrino-blazars association is not found in every analysis, depending on the handling of directional uncertainties \citep[see discussion in][]{2023ApJ...954...75A,OVRO_IC_24}. Utilizing additional observables is thus crucial to refine and potentially confirm the connection.

With the growing sample of neutrino-coincident blazars, we can investigate the nature of the jet-related effects on the neutrino emission in more detail. In this paper, we focus on a direct study of the jet geometry and kinematics of blazars that are likely neutrino sources. Specifically, we explore beaming indicators, such as the Doppler factor and the jet viewing angle, for blazars that coincide with high-energy neutrinos detected at the IceCube (\autoref{s:data}). This comparison enables us to establish more directly that neutrinos are effectively beamed, similar to electromagnetic emission (\autoref{s:beaming-obs}). We discuss the implications of these findings for constraining the neutrino production region and the jet acceleration mechanisms (\autoref{s:beaming-interpret}, \autoref{s:conclusion}). We demonstrate the robustness of the findings presented in this paper and of our earlier results by the consistency between different IceCube event samples in \autoref{s:significance} and \autoref{s:beaming-obs}.

\section{Data}
\label{s:data}

\begin{figure*}
\centering
\includegraphics[width=0.95\linewidth,trim={0.8cm 1.1cm 0.8cm 1.3cm},clip]{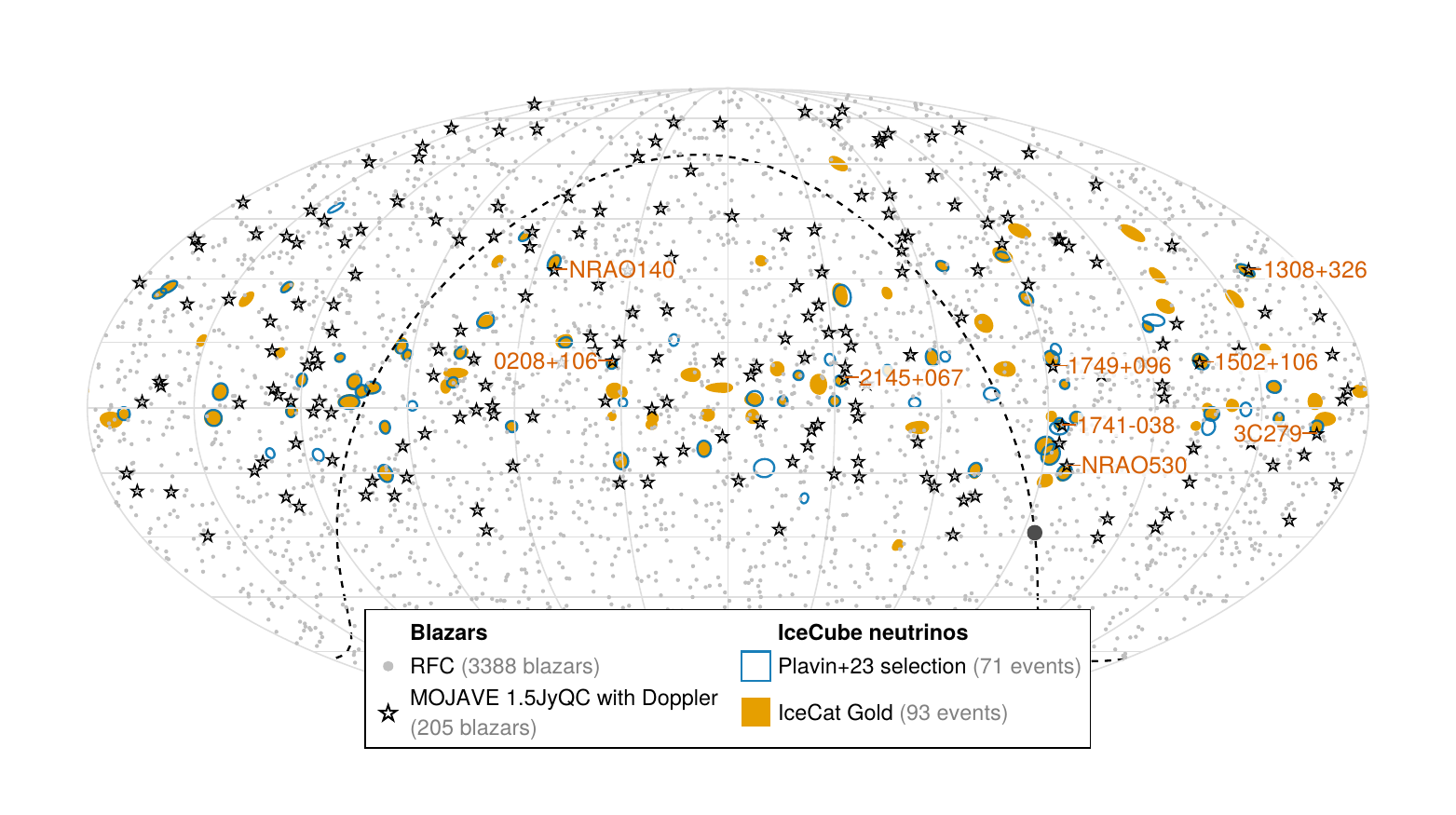}
\caption{
The sky distribution of radio blazars and IceCube events used in this work. See \autoref{s:data} for the definitions of each sample. Two different IceCube event selections are shown in different manners (filled and unfilled) to enable a clearer visual comparison between them. There are 83 blazars from the RFC catalog coincident with at least one IceCube neutrino; note that they include both real physical associations and chance coincidences, see \autoref{f:vlbi_flux_nu}. Neutrino-coincident MOJAVE 1.5JyQC (\autoref{s:data_vlbi}) blazars are labeled here; see \autoref{f:images} for their VLBI images and \autoref{tab:mojmatches} for their properties.
\label{f:skymap}
}
\end{figure*}

\subsection{Radio observations of parsec-scale emission}
\label{s:data_vlbi}

In this work, we follow the strategy of earlier studies of the neutrino-blazar connection (e.g., \citealt{Plavin20} and \citealt{Plavin23} referred hereafter as \citetalias{Plavin20} and \citetalias{Plavin23}; also, \citealt{OVRO_IC_21,OVRO_IC_24,2024ApJ...964....3A}) and rely on Very Long Baseline Interferometry (VLBI) observations to uniformly select bright blazars across the sky. A complete sample of VLBI-selected blazars, along with their positions and parsec-scale flux densities, is compiled in the Radio Fundamental Catalogue \citep[RFC,][]{RFC,doi.org10.25966dhrk-zh08}. Here, we use the RFC version 2019c, the same as earlier in \citetalias{Plavin20,Plavin23}. VLBI observations defining this sample were performed at the 8~GHz band, and include geodetic VLBI \citep{2009JGeod..83..859P,2012A&A...544A..34P,2012ApJ...758...84P}, the Very Long Baseline Array (VLBA) calibrator surveys \citep[VCS;][]{2002ApJS..141...13B,2003AJ....126.2562F,2005AJ....129.1163P,2006AJ....131.1872P,2007AJ....133.1236K,2008AJ....136..580P,r:wfcs,2016AJ....151..154G}, other 8~GHz global VLBI, VLBA, EVN (the European VLBI Network), and LBA (the Australian Long Baseline Array) observations \citep{2011AJ....142...35P,2011AJ....142..105P,2011MNRAS.414.2528P,2012MNRAS.419.1097P,2013AJ....146....5P,2015ApJS..217....4S,2017ApJS..230...13S,2019MNRAS.485...88P,2021AJ....161...88P}. The complete flux density-limited sample consists of 3370 objects with a historic median 8-GHz flux density $S^\mathrm{VLBI}_\mathrm{8\,GHz} \geq 150$~mJy integrated over their VLBI images; these objects are shown in \autoref{f:skymap}. The typical uncertainty of an individual VLBI flux density measurement is around 10\% \citep[dominated by the amplitude calibration;][]{RFC}.

For Doppler boosting and jet viewing angle measurements, we rely on the results of the MOJAVE (Monitoring Of Jets in Active galactic nuclei with VLBA Experiments) team. These parameters cannot be measured directly for blazars; the MOJAVE team has suggested \citep{Homan06} and applied \citep{MOJAVEdoppler} a procedure to infer them from VLBI monitoring data; we use those measurements in this paper. The full MOJAVE dataset is assembled with a range of selection criteria: bright VLBI-selected blazars, compact symmetric objects, Fermi-selected blazars, low-luminosity AGNs \citep{FM1,MOJAVEangles,2020ApJ...899..141L,2013AJ....146..120L}. Similarly to our earlier studies, we focus on a uniform flux-density limited subsample of the MOJAVE program sources in the analysis: the 1.5~Jy Quarter Century Sample (1.5JyQC; defined in \citealt{MOJAVEangles}). It consists of blazars at $\delta > -30\degree$ with VLBI flux density exceeding 1.5~Jy at 15~GHz at least once during the quarter century of observations, 1994--2019. This sample contains 232~blazars, and 205 of those have Doppler boosting measurements (highlighted in \autoref{f:skymap}). The MOJAVE 1.5JyQC sample was specifically designed to be homogeneous and unbiased.

\subsection{IceCube high-energy neutrinos}
\label{s:data_icecube}

From the neutrino side, we utilize the high-energy events reported by the IceCube observatory. We consider two high-energy event catalogs, both to ensure the consistency of our results and keep the continuity with earlier studies. Both datasets are analyzed in the exact same way, and we ensure that all reported results are consistent between the two.

\paragraph*{The P23 IceCube event selection.}~
The first catalog is the sample of high-energy alert-like neutrino detections we collected before, most recently in \citetalias{Plavin23}. Earlier, the same selection criteria were utilized in \citetalias{Plavin20} and \citet{OVRO_IC_21}. This sample consists of 71 events with the angular uncertainty region below 10~sq.\,deg.\ and the estimated energy above 200~TeV.

\paragraph*{The IceCat-1 event catalog.}~
In 2023, the IceCube collaboration released the IceCat-1 track event catalog \citep[][last updated in November 2023]{DVN/SCRUCD_2023}. It is mostly based on a different reconstruction of IceCube detection events, utilizes different event selection and categorization criteria. Following the IceCube guidance on event categories in the catalog, we select 93~Gold events from the IceCat-1 with the angular uncertainty region below 10~sq.\,deg. The estimated signalness is $\geq50\%$ for all these events; their neutrino energy estimates range from 56~TeV to 6~PeV.

The sky localization of IceCube events from both samples is shown in \autoref{f:skymap}. Blazars coincident with IceCube events are highlighted in the figure, and those from the MOJAVE 1.5JyQC sample are labeled. See \autoref{tab:mojmatches} for the properties of neutrino-coincident blazars from the full MOJAVE sample. Note how the change of IceCube event parameters between different reconstructions leads to slightly different lists of events and of their blazar coincidences. Evaluating the reconstruction differences in details is out of scope of this paper; here, we use these two catalogs as an extra robustness check of our results, reporting only conclusions consistent between the two.

For the analyses performed in this work, we enlarge the original 90\% directional uncertainties reported in IceCube catalogs by $0.78\degree$ isotropically in all directions. This enlargement magnitude was determined in \citetalias{Plavin23} to be optimal for association searches, and we keep the same value to avoid multiple trials in the statistical analysis. As a guidance, \autoref{tab:mojmatches} marks the three coincidences that don't require this enlargement. A similar enlargement was found in \citet{OVRO_IC_24}; see that paper, \citetalias{Plavin23}, and \autoref{s:significance} for a more detailed discussion.

\section{Neutrino-blazar correlation}
\label{s:significance}

\begin{deluxetable*}{ccccccccccc}
\tablecaption{Key properties of high-energy neutrino coincident blazars that are observed within the MOJAVE program. \label{tab:mojmatches}}
\tablehead{
\multicolumn{2}{c}{Neutrino event} & \multicolumn{2}{c}{Blazar name} & \colhead{} & \multicolumn{2}{c}{VLBI Flux} & \colhead{} & \multicolumn{3}{c}{Beaming Indicators} \\
\cline{1-2}
\cline{3-4}
\cline{6-7}
\cline{9-11}
\colhead{Date} & \colhead{Event ID} & \colhead{J2000} & \colhead{B1950} & \colhead{$z$} &
\colhead{8 GHz} & \colhead{15 GHz} & \colhead{1.5JyQC} & \colhead{$D$} & \colhead{$\Gamma$} & \colhead{$\theta$} \\
\colhead{(P23 sample)} & \colhead{(IceCat Gold sample)} & & \colhead{} & \colhead{} & \colhead{(Jy)} & \colhead{(Jy)} & \colhead{} & \colhead{} & \colhead{} & \colhead{($^{\circ}$)}\\
\colhead{(1)} & \colhead{(2)} & \colhead{(3)} & \colhead{(4)} & \colhead{(5)} & \colhead{(6)} & \colhead{(7)} & \colhead{(8)} & \colhead{(9)} & \colhead{(10)} & \colhead{(11)}
}
\startdata
2015$-$09$-$26\textsuperscript{b} & IC150926A\textsuperscript{b} & J1256$-$0547 & 1253$-$055 & $ 0.536$ & $   15$ & $   19$ &  yes & $  140$ & $   72$ & $ 0.12$\\ 
2022$-$03$-$03\textsuperscript{b} &  & J1751$+$0939 & 1749$+$096 & $ 0.322$ & $  2.8$ & $    4$ &  yes & $  100$ & $   51$ & $ 0.07$\\ 
2019$-$07$-$30\textsuperscript{a} & IC190730A\textsuperscript{a} & J1504$+$1029 & 1502$+$106 & $ 1.838$ & $  1.5$ & $  1.4$ &  yes & $   65$ & $   35$ & $ 0.44$\\ 
2016$-$01$-$28\textsuperscript{b} & IC160128A\textsuperscript{b} & J1733$-$1304 & 1730$-$130 & $ 0.902$ & $    4$ & $  4.5$ &  yes & $   48$ & $   32$ & $    1$\\ 
2015$-$08$-$12\textsuperscript{b} & IC150812B\textsuperscript{b} & J2148$+$0657 & 2145$+$067 & $ 0.999$ & $  6.6$ & $  7.5$ &  yes & $   31$ & $   16$ & $ 0.36$\\ 
2015$-$08$-$31\textsuperscript{b} & IC150831A\textsuperscript{b} & J0336$+$3218 & 0333$+$321 & $ 1.259$ & $  1.6$ & $  1.7$ &  yes & $   30$ & $   18$ & $  1.4$\\ 
2012$-$05$-$15\textsuperscript{b} & IC120515A\textsuperscript{b} & J1310$+$3220 & 1308$+$326 & $ 0.997$ & $  1.9$ & $  2.1$ &  yes & $   24$ & $   28$ & $  2.3$\\ 
2022$-$02$-$05\textsuperscript{b} & IC220205B\textsuperscript{b} & J1743$-$0350 & 1741$-$038 & $ 1.054$ & $    4$ & $  5.3$ &  yes & $   22$ & $     $ & $     $\\ 
2011$-$09$-$30\textsuperscript{b} &  & J1743$-$0350 & 1741$-$038 & $ 1.054$ & $    4$ & $  5.3$ &  yes & $   22$ & $     $ & $     $\\ 
2013$-$10$-$14\textsuperscript{b} & IC131014A\textsuperscript{b} & J0211$+$1051 & 0208$+$106 & $   0.2$ & $ 0.66$ & $  1.1$ &  yes & $  6.3$ & $  5.1$ & $  8.8$\\ 
2020$-$11$-$30\textsuperscript{a} & IC201130A\textsuperscript{a} & J0206$-$1150 & 0203$-$120 & $ 1.663$ & $ 0.28$ & $  0.3$ &   no & $   35$ & $   20$ & $    1$\\ 
2017$-$09$-$22\textsuperscript{b} & IC170922A\textsuperscript{b} & J0505$+$0459 & 0502$+$049 & $ 0.954$ & $  0.7$ & $ 0.93$ &   no & $   25$ & $   14$ & $  1.6$\\ 
 & IC221223A\textsuperscript{b} & J2311$+$3425 & 2308$+$341 & $ 1.817$ & $ 0.69$ & $ 0.86$ &   no & $   23$ & $     $ & $     $\\ 
 & IC201221A\textsuperscript{b} & J1724$+$4004 & 1722$+$401 & $ 1.049$ & $  0.5$ & $ 0.58$ &   no & $  9.8$ & $   61$ & $  3.2$\\ 
2010$-$10$-$09\textsuperscript{b} &  & J2200$+$1030 & 2157$+$102 & $ 0.172$ & $ 0.19$ & $ 0.22$ &   no & $    5$ & $     $ & $     $\\ 
2017$-$09$-$22\textsuperscript{a} & IC170922A\textsuperscript{a} & J0509$+$0541 & 0506$+$056 & $0.3365$ & $ 0.42$ & $ 0.46$ &   no & $  1.8$ & $  1.5$ & $   33$\\ 
2011$-$07$-$14\textsuperscript{b} & IC110714A\textsuperscript{b} & J0432$+$4138 & 0429$+$415 & $ 1.022$ & $  1.4$ & $    1$ &   no & $ 0.13$ & $   54$ & $   31$\\
\enddata
\tablecomments{Columns are as follows: (1)~---~IceCube neutrino arrival date for events included in the \citetalias{Plavin23} sample; (2)~---~IceCat event ID for events included in IceCat-1; see \autoref{s:data_icecube} for a definition of these samples; \textsuperscript{a} indicates neutrino-blazar matches within original reported uncertainties, and \textsuperscript{b} those matches that require extending original uncertainties, following \autoref{s:significance}.\\ (3), (4)~---~coordinate-based blazar names, with the J2000 name taken directly from the RFC \citep{RFC}; (5)~---~redshift $z$ extracted from the NASA/IPAC Extragalactic Database (NED); (6), (7)~---~parsec-scale radio flux densities measured by VLBI, as presented in the RFC catalog; their typical uncertainty is about $10\%$, dominated by the amplitude calibration; (8)~---~flag indicating whether the blazar is included in the uniformly-selected MOJAVE 1.5JyQC sample (\autoref{s:data_vlbi}); those sources are also labeled in \autoref{f:skymap}, and their VLBI images are shown in \autoref{f:images}; (9), (10), (11)~---~beaming indicators, Doppler $D$ and Lorentz $\Gamma$ factors and the viewing angle $\theta$, as measured within the MOJAVE program \citep{MOJAVEdoppler}; see \autoref{s:beaming-obs} for a discussion of assumptions and uncertainties of these measurements.\\
The blazars are ordered by the inclusion in MOJAVE 1.5JyQC (8) and by the Doppler factor measurement (9).}
\end{deluxetable*}

The statistical correlation between VLBI-bright blazars and high-energy neutrinos has already been established and, as confirmed in our previous work, strengthens over time with more neutrino detections (most recently, \citetalias{Plavin23}). In this work, we do not aim to repeat or reevaluate that analysis. Instead, we utilize another dataset of IceCube events (IceCat-1, \autoref{s:data_icecube}) to serve as an additional check of the robustness of our findings. Both the IceCat-1 dataset and the neutrino events sample from \citetalias{Plavin23} are used in the following sections of this paper.

We briefly remind the reader of the statistical testing procedure. We evaluate the presence and statistical significance of the correlation between high-energy neutrinos and VLBI bright blazars in the same way as in \citetalias{Plavin20,Plavin23}. First, we choose a test statistic: the logarithmic (geometric) average of VLBI fluxes of all blazars from \autoref{s:data_vlbi} falling within IceCube event error ellipses as defined in \autoref{s:data_icecube}. The mean of the logarithms is preferrable to the arithmetic mean here because measured fluxes cover several orders of magnitude, and relative differences between them are physically important. Statistically, we expect that using the arithmetic mean would lead to a lower but comparable test power.

After having chosen the test statistic, we calculate the probability that its observed value could have arisen by chance, assuming no directional neutrino-blazar correlation. For that, we randomize the Right Ascensions of all IceCube events $10^6$~times, motivated by the instrument sensitivity being dependent on the Declination only \citep[used in][and elsewhere]{2017ApJ...835..151A}. Then, the chance probability is the fraction of random realizations where the average VLBI flux of matching blazars is higher than the observed value. There are no free parameters to be optimized in this procedure, and we can immediately interpret the resulting fraction as the $p$-value.

\begin{figure*}
\centering
\includegraphics[width=0.48\linewidth]{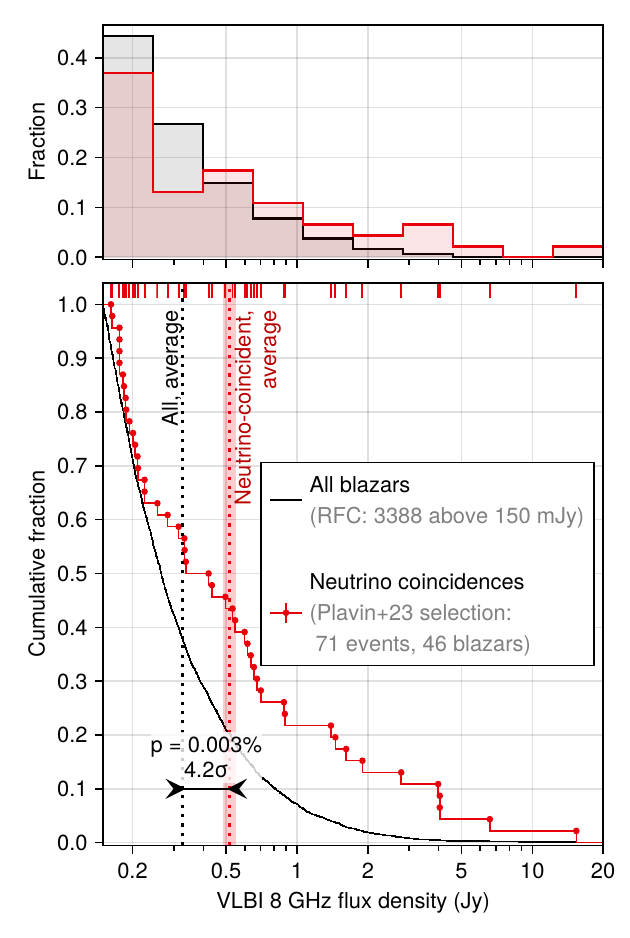}
\includegraphics[width=0.48\linewidth]{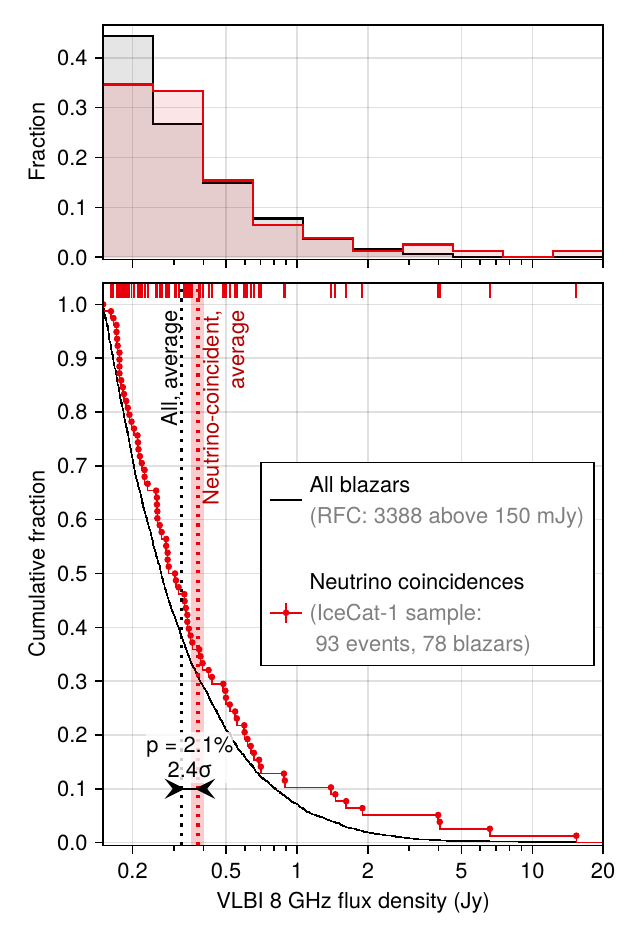}
\caption{
    The distribution of VLBI 8~GHz flux density for blazars coincident with IceCube neutrinos, shown as histograms and cumulative distribution plots. Similarly to \autoref{f:skymap}, all VLBI blazars from the RFC catalog are shown in black exactly the same in the left and right panels. Neutrino coincidences are shown in red, using the \citetalias{Plavin23} events sample on the left and the IceCat-1 on the right (see \autoref{s:data_icecube}).
}
\label{f:vlbi_flux_nu}
\end{figure*}

\paragraph*{The P23 IceCube event selection.}~
The VLBI blazar-neutrino correlation for this event sample was performed in \citetalias{Plavin23}, resulting in a $3.6\sigma$ significance ($4.2\sigma$ pre-trial). For ease of comparison and to remind the reader, we visualize this result in \autoref{f:vlbi_flux_nu}.

\paragraph*{The IceCat-1 event catalog.}~
To validate our earlier findings, we also utilize the IceCat-1 Gold dataset released by IceCube in 2023 (\autoref{s:data_icecube}). Applying the same statistical procedure, we obtain the chance probability of $p=2.1\%$ ($2.4\sigma$; no trials) as shown in \autoref{f:vlbi_flux_nu}. The correlation remains convincing, albeit with a significance level lower than for our previous event selection. We find such an agreement between results obtained from different event reconstructions, selections, and categorization approaches quite remarkable.

Note that updated or more detailed event reconstructions are not necessarily better in every aspect: see demonstrations and discussion in the review \cite{2023arXiv231100281T} and in \cite{2024arXiv240303752S}. For example, a reconstruction update lead to a decrease of the TXS~0506+056 flare significance \citep{2021arXiv210109836I,2023arXiv230714559L}. A detailed analysis and interpretation of the differences between various IceCube event reconstructions is out of scope of this work.

Overall, we see evidence for neutrino-blazar correlation using either the \citetalias{Plavin23} or the IceCat-1 samples. Both give consistent results and are used in the following sections of this paper for more detailed studies. This helps us ensure that we only report findings that hold across different neutrino event samples.

\paragraph*{Cross-check without enlarging IceCube errors.}~
As detailed in \autoref{s:data_icecube}, we enlarge original IceCube event uncertainties by $0.78\degree$ for all calculations in this paper, following the approach of \citetalias{Plavin23} and \citet{OVRO_IC_24} and avoiding multiple comparisons issues. Here, we present a focused cross-check that such an enlargement remains important for both IceCube event catalogs we use (\autoref{s:data_icecube}).

Repeating our statistical testing procedure with original IceCube uncertainties taken as-is, we don't see any significant evidence for the correlation with VLBI blazars. The chance probability $p > 5\%$ for both IceCube event selections. The spatial correlation shown above is heavily driven by blazars that lie slightly outside the original reported 90\% directional errors. This is still consistent with the observed correlation caused by a physical connection, see \cite{2025A&A...696A..73K} for a sensitivity and robustness analysis.

Such an effect could result from unaccounted systematic uncertainties or other discrepancies between estimated and actual error regions. A detailed dedicated study can only be performed by the IceCube collaboration, utilizing all the available data about each detection; see \cite{2021arXiv210708670L} for a discussion and an overview of challenges. Further in the future, it will be crucial to compare the brightest neutrino sources in the sky between different neutrino observatories.
The cross-check we performed here does not replace the detailed analysis performed in \citetalias{Plavin23} and \cite{OVRO_IC_24}, but remains consistent with those works and results reported by \cite{2023ApJ...954...75A,2024ApJ...973...97A,2023ApJ...955L..32B,2025arXiv250304632K}.

\section{Geometry and kinematics of neutrino-associated blazars}
\label{s:beaming}

\begin{figure*}
\centering
\includegraphics[width=0.98\linewidth]{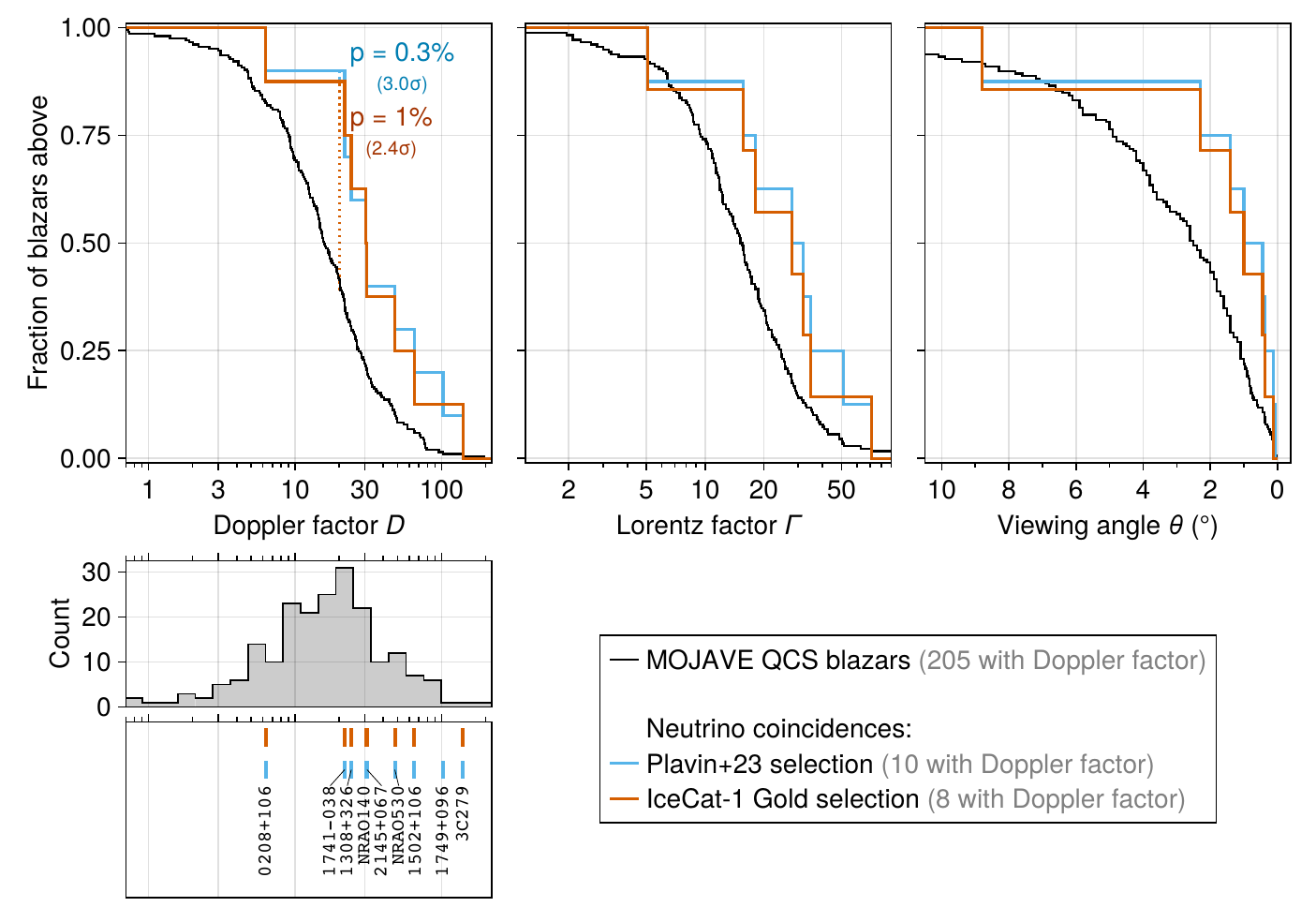}
\caption{
Comparison of observational jet beaming indicators (Doppler $D$ and Lorentz $\Gamma$ 
factors, viewing angle $\theta$) for neutrino-coincident blazars and for the complete MOJAVE 1.5JyQC sample (see \autoref{s:data} and \autoref{tab:mojmatches}).
Shown are cumulative distributions for all three parameters (the top row). The Doppler factor is among the most direct observational tracers, and we show its histogram with individual neutrino coincidences marked below. Note: for the \citetalias{Plavin23} neutrino sample, there are 9 coincident blazars; one of them is coincident with two neutrinos, adding up to 10 coincidences. The chance probabilities to obtain the observed Doppler factor excess for neutrino coincidences are shown in the plot (see \autoref{s:beaming} for details).
\label{f:nudoppler}
}
\end{figure*}

\subsection{Observational evidence for neutrino beaming}
\label{s:beaming-obs}

The association of high-energy neutrinos with VLBI-bright blazars (\autoref{s:significance}) already suggests that beaming plays a major role in neutrino production and detection: neutrinos are predominantly emitted along the direction of the jet. Indeed, VLBI-selected AGNs are dominated by galaxies with jets pointing towards us, their typical viewing angles are $\theta = 2\dots3\degree$ \citep{MOJAVEangles,MOJAVEdoppler}. If neutrino emission were isotropic, we would expect to see thousands ($4\pi/\pi\theta^2~\sim~3000$) of similar AGNs pointed in other directions for each neutrino-VLBI association. This is clearly incompatible with the observed number of high-energy neutrinos.

To further probe and quantify the beaming of neutrino emission more directly, we explore kinematic and geometric properties, measured for individual blazars. Specifically, we analyze the plasma Lorentz factor $\Gamma$ (the bulk jet velocity), the viewing angle $\theta$ (between the jet direction and the Earth), and the Doppler factor $D$ (a key measure of relativistic beaming).
These three quantities characterize the plasma flow geometry in different but related ways:
\begin{equation}
\begin{split}
\Gamma &= \frac{1}{\sqrt{1 - \beta^2}}\\
D &= \frac{1}{\Gamma  (1 - \beta \cos\theta)}
\end{split}
\end{equation}
They are measured from VLBI observations within the MOJAVE program (\autoref{s:data_vlbi}, \citealt{MOJAVEdoppler}). These estimates crucially rely on the VLBI ability to resolve the parsec-scale structure and determine the brightness of the most compact detail~--~the so-called core. Due to the measurement procedure, the estimates of these quantities are somewhat related to the VLBI flux density we use in the correlation analysis above (\autoref{s:significance}). Still, they are not exactly the same and differ in fundamental ways. The VLBI flux density is a measure of the total radio emission on multi-parsec scale, while Doppler factors are determined from the brightness of even more compact regions, typically a fraction of a parsec in projection. Viewing angle and Lorentz factor estimates additionally rely on parsec-scale kinematics to constrain the geometry. It can be challenging to calculate precise uncertainties of these measurements. Still, as a cross-check, Doppler factors resulting from MOJAVE measurements were found to be consistent with independent estimates obtained via different techniques, such as radio variability \citep{2018ApJ...866..137L,MOJAVEdoppler}.

We probe the jet beaming --- neutrino emission connection by comparing these beaming indicators (Doppler factor, plasma velocity, viewing angle) between neutrino-coincident blazars and the overall sample. To ensure maximal robustness of results, we remain conservative and focus on a uniform flux density-limited MOJAVE subsample, 1.5JyQC (\autoref{s:data_vlbi}).
The comparison of blazars close to neutrino arrival directions with the whole MOJAVE 1.5JyQC is shown in \autoref{f:nudoppler}. All the beaming indicators demonstrate a consistent picture: blazars consistent with neutrino directions have higher Doppler and Lorentz factors, and smaller viewing angles. All of these indicate extremely highly beamed jets, even higher than typical VLBI-selected blazars. A to-scale comparison is shown in \autoref{f:geometry}.

We evaluate the statistical significance of the Doppler factor correlation with neutrino coincidences. Doppler factors are chosen because they are the most direct tracers of the electromagnetic emission beaming. They are also measured more directly from observations, see \cite{MOJAVEdoppler}. We rely on the same statistical testing procedure as in \autoref{s:significance}, namely the shuffling of neutrino event positions in the sky. We choose the Kolmogorov-Smirnov test statistic between the Doppler factor distributions for MOJAVE 1.5JyQC blazars and for neutrino coincidences.

This test is visualized in \autoref{f:nudoppler}; it demonstrates that the apparent differences are indeed unlikely to arise by chance. The chance probability to obtain the observed correlation is $p=0.3\%$ ($3.0\sigma$) for the \citetalias{Plavin23} event selection, and $p=1\%$ ($2.4\sigma$) for IceCat-1 Gold. We stress that this analysis focuses on the complete VLBI-flux-density limited MOJAVE sample, 1.5JyQC. As in earlier studies, here we chose the cleanest uniform sample of sources for the primary analysis to remain conservative, to keep statistical purity and robustness.
Simply extending it to the full MOJAVE dataset would increase the significance level to $3.5\sigma$, see \autoref{s:fullmojave}. 
Unlike the 1.5JyQC sample, the full MOJAVE dataset contains sources selected by different criteria accumulated since the start of the program in 1994, see \autoref{s:data_vlbi}.

The most extreme viewing angle estimates from the MOJAVE observations can be less than $0.1\degree$. There are only three sources with such viewing angles in the full sample \citep{MOJAVEdoppler}, and one of them is coincident with an IceCube neutrino: 1749+096 with $\theta=0.07\degree$, see \autoref{tab:mojmatches} and \autoref{f:images}. We believe that such blazars do indeed have jets pointing almost directly towards us, but recommend treating specific values of these extreme viewing angle estimates with caution. It is possible that some limitations of the generic approach utilized in \cite{MOJAVEdoppler} manifest in such extreme cases. Additional or alternative ways to infer the jet geometry and the beaming patterns would potentially be useful for more detailed quantitative studies in the future. For example, a very large apparent jet opening angle \citep{2017MNRAS.468.4992P} might be used as another independent indicator of a very small viewing angle (e.g., the case of PKS~1424+240, \citealt{2025arXiv250409287K}).

Blazars with activity patterns wildly varying over time may also hit limitations of the uniform MOJAVE analysis. In particular, estimates for TXS~0506+056 presented in \autoref{tab:mojmatches} are dominated by its behavior before the major flare in 2018-2022 that followed the neutrino detection. Before that, it was not bright enough to be included into the 1.5JyQC sample, and the lack of activity resulted in low Doppler factor and large viewing angle estimates. Such cases may deserve a dedicated study that takes the flaring behavior into account.

\subsection{Beamed neutrino emission implications}
\label{s:beaming-interpret}

\begin{figure*}
\centering
\includegraphics[width=0.99\linewidth]{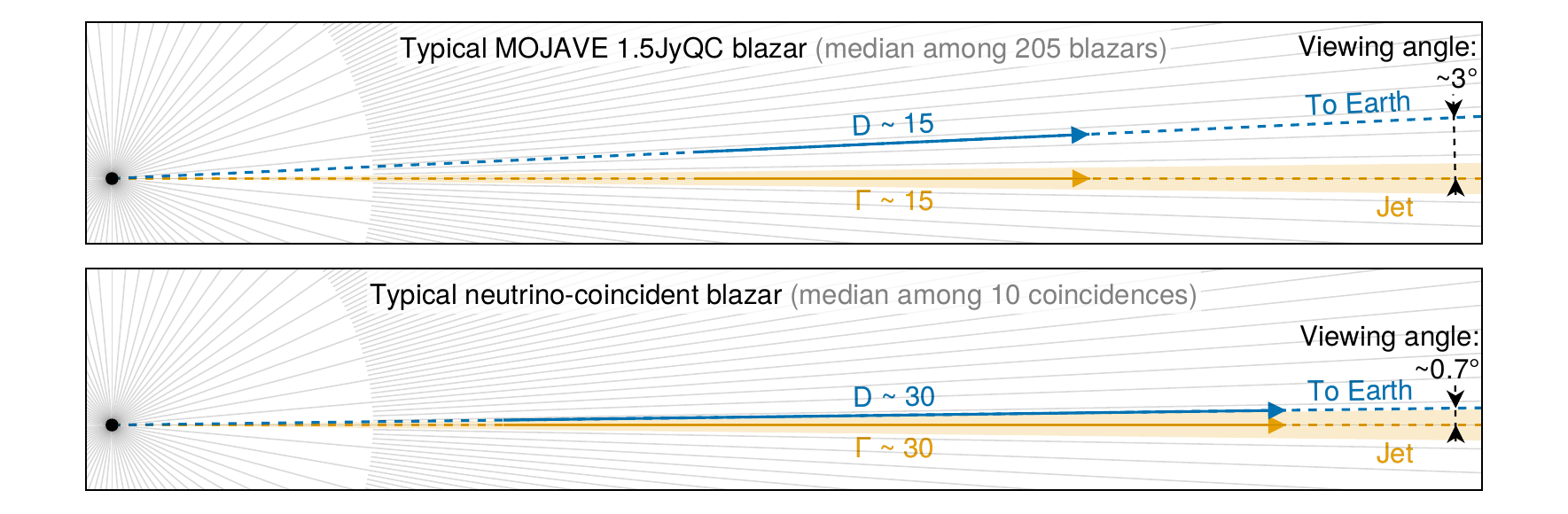}
\caption{
Schematic illustration comparing the typical geometries for blazars in the MOJAVE 1.5JyQC sample (top; \citealt{MOJAVEdoppler}) and neutrino-coincident blazars (bottom; \autoref{tab:mojmatches}). Angles are shown to scale, with gray lines indicating $1\degree$ and $5\degree$ increments for guidance. All quantities depicted are medians within their respective samples. The visual scaling of velocity and beaming factors is arbitrary.
\label{f:geometry}
}
\end{figure*}

Findings presented in \autoref{s:significance} and \autoref{s:beaming-obs} provide an evidence for a high beaming of the jet radio emission in neutrino-associated blazars at a range of scales. From a fraction of a parsec (VLBI core brightness, Doppler factors) to tens of parsecs (total VLBI flux density, kinematics measurements), we see that blazars coincident with neutrino events are even more extreme than other VLBI blazars (\autoref{f:geometry}). This unambiguously implies that the neutrino emission itself is highly anisotropic, these particles are predominantly emitted along the jet axis. They are detected on Earth preferentially from sources with the jet pointing straight towards us. The beaming appears to affect both neutrinos and photons in a similar way, but detailed quantification of this effect requires a more reliable identification of individual neutrino-emitting blazars and their properties.

At production, the directionality of emitted neutrinos is mostly governed by the parent high-energy proton momentum, in both $pp$ and $p\gamma$ scenarios. The predominance of neutrinos along the jet axis indicates that these protons have a significant bulk motion in the jet in the region of neutrino production. If electrons and protons gain their bulk acceleration at similar spatial scales, then the neutrino production happens where the jet is already accelerated. This typically occurs at distances of about a parsec or more from the jet origin \citep[e.g.,][]{2012ApJ...745L..28A,2015ApJ...798..134H,2020MNRAS.495.3576K}. Alternatively, high-energy protons may be efficiently accelerated much closer to the black hole than most of the electrons \citep[e.g.][]{PtitsynaNeronov}. Distinguishing between these scenarios is important to reliably constrain the proton interaction and neutrino production region. Potentially, this would lead to a lower limit on the distance from the black hole, nicely complementing the upper limit of $\lesssim$~parsecs arising from VLBI (\citetalias{Plavin20,Plavin23}; \autoref{s:significance}) and temporal  (e.g.~ \citetalias{Plavin20}; \citealt{OVRO_IC_21,2023MNRAS.519.1396S,OVRO_IC_24,2024ApJ...964....3A,2024MNRAS.527.8784A}) correlations.

Neutrino production at $\sim$~parsec distances from the central black hole is consistent with the likely scenario of $p\gamma$ interactions within the jet.
There, the required X-ray target photon field can be produced either by the jet itself \citep[self- or external Compton,][]{neutradio2,Polina,2025arXiv250201738F} or by other regions of the AGN, such as the corona \citep[e.g.,][]{Inoue-corona,Murase-corona,Simon-corona}.
The exact origin of these photons remains an open question, together with the specific mechanism accelerating protons to PeV energies.

We note that our findings do not preclude the presence of other, non-beamed, sources of neutrinos. The neutrino--VLBI flux density and neutrino--Doppler factor correlations imply that there is a population of blazars that emit neutrinos, and emit them predominantly along the jet axis. Non-beamed and non-jet-related neutrino production, such as possible for Seyfert-type AGNs like NGC1068 \citep{IceCubeNGC1068,2024arXiv240607601A,NeronovSeyferts}, may form another population of sources leading to a diverse extragalactic neutrino sky.

Heavily flaring blazars producing neutrino during their flares may potentially form another population of neutrino sources that our analysis is not sensitive to. Earlier studies reported observational evidence for neutrinos coming preferentially during electromagnetic flux increases, especially in the radio band: \citetalias{Plavin20}; \cite{2024MNRAS.527.8784A,OVRO_IC_21,2024ApJ...964....3A,OVRO_IC_24}. Those sources may or may not demonstrate extreme jet beaming and persistently bright radio emission. For example, TXS~0506+056 exhibited barely any activity before the neutrino arrival and the following major radio flare in 2018~--~2022. This likely affects its beaming estimates, see \autoref{s:beaming-obs}. Meanwhile, PKS~1502+106 is another heavily flaring blazar that was originally selected for its highest temporal neutrino-radio correlation in \citetalias{Plavin23} and later studied as a likely neutrino counterpart \citep[e.g.][]{2021JCAP...10..082O,2021MNRAS.503.3145B}. Unlike TXS~0506+056, this blazar demonstrates an extremely strong beaming in our analysis, see \autoref{tab:mojmatches}.

\section{Summary}
\label{s:conclusion}

In this work, we studied direct observational tracers of jet beaming in blazars coincident with high-energy neutrinos detected with the IceCube. The blazars located around neutrino arrival directions were found to exhibit extremely high beaming with median Doppler factor of $D \approx 30$ and a median viewing angle of $\theta \approx 0.7\degree$. These blazars are even more extreme than already highly beamed VLBI-selected ones that typically exhibit $D \approx 15$, $\theta \approx 3 \degree$ (\autoref{f:geometry}). The statistical significance of the Doppler factor versus neutrino correlation is $3\sigma$ and $2.4\sigma$ for different IceCube event selections (see \autoref{s:beaming} and \autoref{f:nudoppler}).

This population study provides yet another piece of evidence that bright blazars are strong neutrino emitters, and the observed neutrino-radio blazar correlation represents a genuine astrophysical connection, not a statistical fluctuation. The effect is consistent across different IceCube event selections and manifests itself in all observational indicators of the jet beaming as shown in \autoref{f:vlbi_flux_nu} and \autoref{f:nudoppler}. In agreement with earlier studies, enlarging the reported IceCube directional uncertainties (by a previously determined value of $0.78\degree$) is necessary to unravel this connection, see \autoref{s:significance}.

The observational connection between the jet electromagnetic emission beaming and detected neutrinos requires the neutrino emission to be highly beamed as well. The beaming effects appear to be comparable for electromagnetic and neutrino emission, but quantifying them would require both more detailed modeling (to handle observational biases and the intrinsic diversity of blazars) and robust individual neutrino associations. Neutrino beaming implies that the multi-PeV protons producing them exhibit a strong bulk motion along the jet direction. In typical jet acceleration profiles, this motion is only attained at parsec scales, potentially providing a lower limit on the distance from the central black hole to the neutrino production region.

Evaluating direct jet beaming indicators of neutrino-coincident blazars, such as the Doppler factor, already draws a consistent picture of their jets as extremely relativistic and aligned with the line of sight.
This effect is demonstrated here by population analysis, and complemented by dedicated studies of individual sources. In particular, see \citet{2025arXiv250409287K} that finds and explores the extreme jet beaming in the neutrino-associated TeV blazar with low VLBI apparent speed PKS~1424+240 \citep{IceCubeNGC1068}.
Note that these beaming indicators are challenging to measure and can only be determined for a limited subset of blazars with great observational coverage. Directly observable properties such as the parsec-scale emission strength measured by VLBI remain well-suited for larger scale searches for neutrino~--~blazar associations.

\section*{Acknowledgements}

We thank Eduardo Ros and an anonymous referee for comments on the manuscript, Markus B{\"o}ttcher, Anna Franckowiak, and Emma Kun for productive discussions.
A.V.P.\ is a postdoctoral fellow at the Black Hole Initiative, which is funded by grants from the John Templeton Foundation (grants 60477, 61479, 62286) and the Gordon and Betty Moore Foundation (grant GBMF-8273). 
Y.Y.K.\ was supported by the MuSES project, which has received funding from the European Union (ERC grant agreement No 101142396). 
%
The work of S.V.T.\ is supported in the framework of the State project ``Science'' by the Ministry of Science and Higher Education of the Russian Federation under the contract 075-15-2024-541.
The views and opinions expressed in this work are those of the authors and do not necessarily reflect the views of these Foundations.

This research has made use of the NASA/IPAC Extragalactic Database, which is funded by the National Aeronautics and Space Administration and operated by the California Institute of Technology.
This research has made use of data from the Radio Fundamental Catalog \citep[][\href{http://doi.org/10.25966/dhrk-zh08}{doi:10.25966/dhrk-zh08}]{RFC}.
This research made use of the data from the MOJAVE database maintained by the MOJAVE team \citep{2018ApJS..234...12L}.

Code to reproduce most of the figures from this manuscript can be found at \url{https://github.com/aplavin/astroplots.jl/tree/master/nubeam25}.


\bibliographystyle{aasjournal}
\bibliography{nubeam}

\appendix
\section{Beaming of neutrino-coincident blazars in the full MOJAVE sample} \label{s:fullmojave}

\begin{figure*}
\figurenum{A1}
\centering
\includegraphics[width=1.00\linewidth]{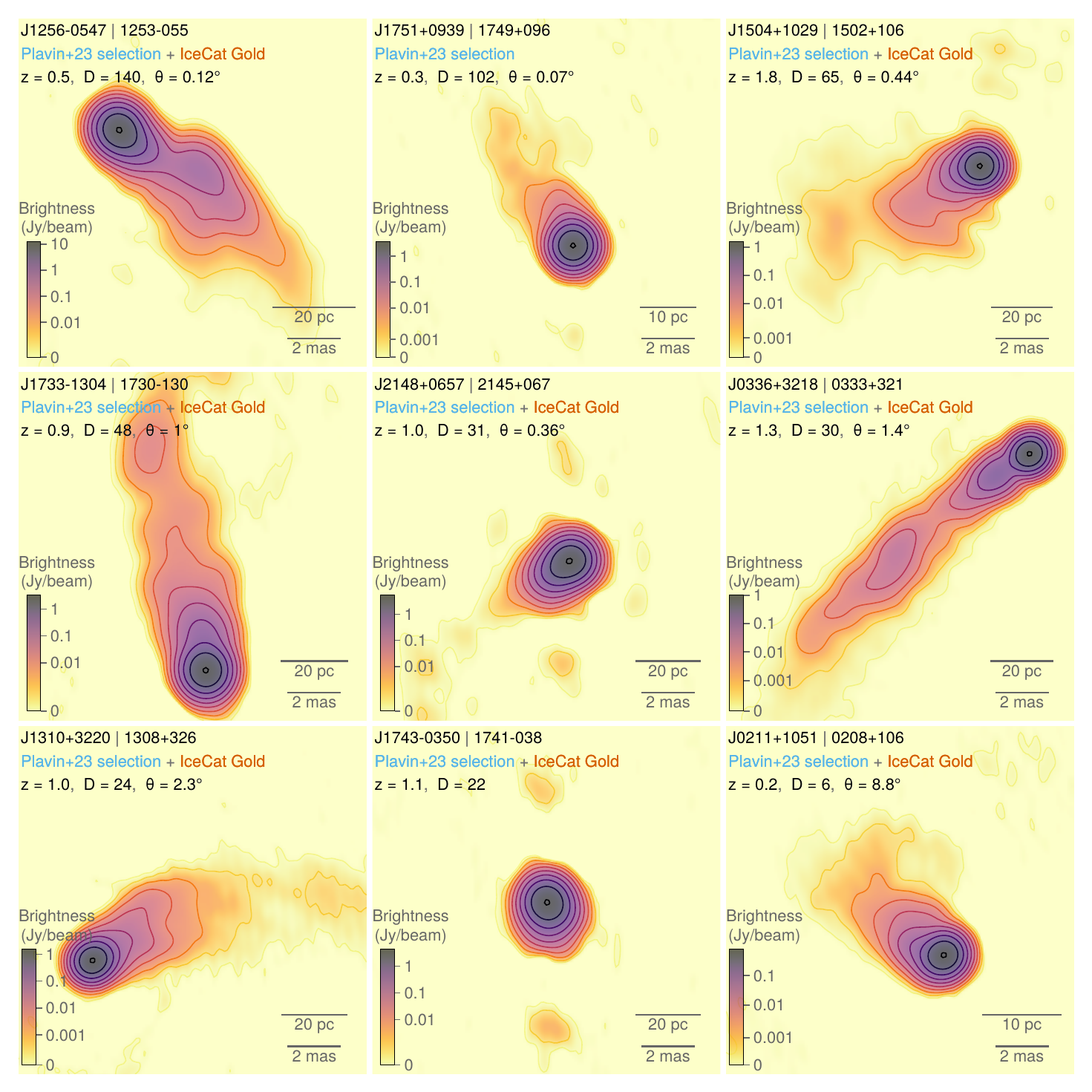}
\caption{
MOJAVE VLBI images for the nine 1.5JyQC blazars coincident with IceCube neutrinos from both samples of \autoref{s:data}. We show here averages of multi-decade individual VLBI images, which provide a better dynamic range to trace the overall structure of the jet. These stacked images are available from the MOJAVE website (\url{https://www.cv.nrao.edu/MOJAVE/}), see details in \citet{2017MNRAS.468.4992P}. For each blazar, we show the redshift $z$, the Doppler factor $D$ and the viewing angle $\theta$ estimates. Blazars are ordered by their Doppler factor, same as in \autoref{tab:mojmatches}.
\label{f:images}
}
\end{figure*}

\begin{figure*}
\figurenum{A2}
\centering
\includegraphics[width=0.98\linewidth]{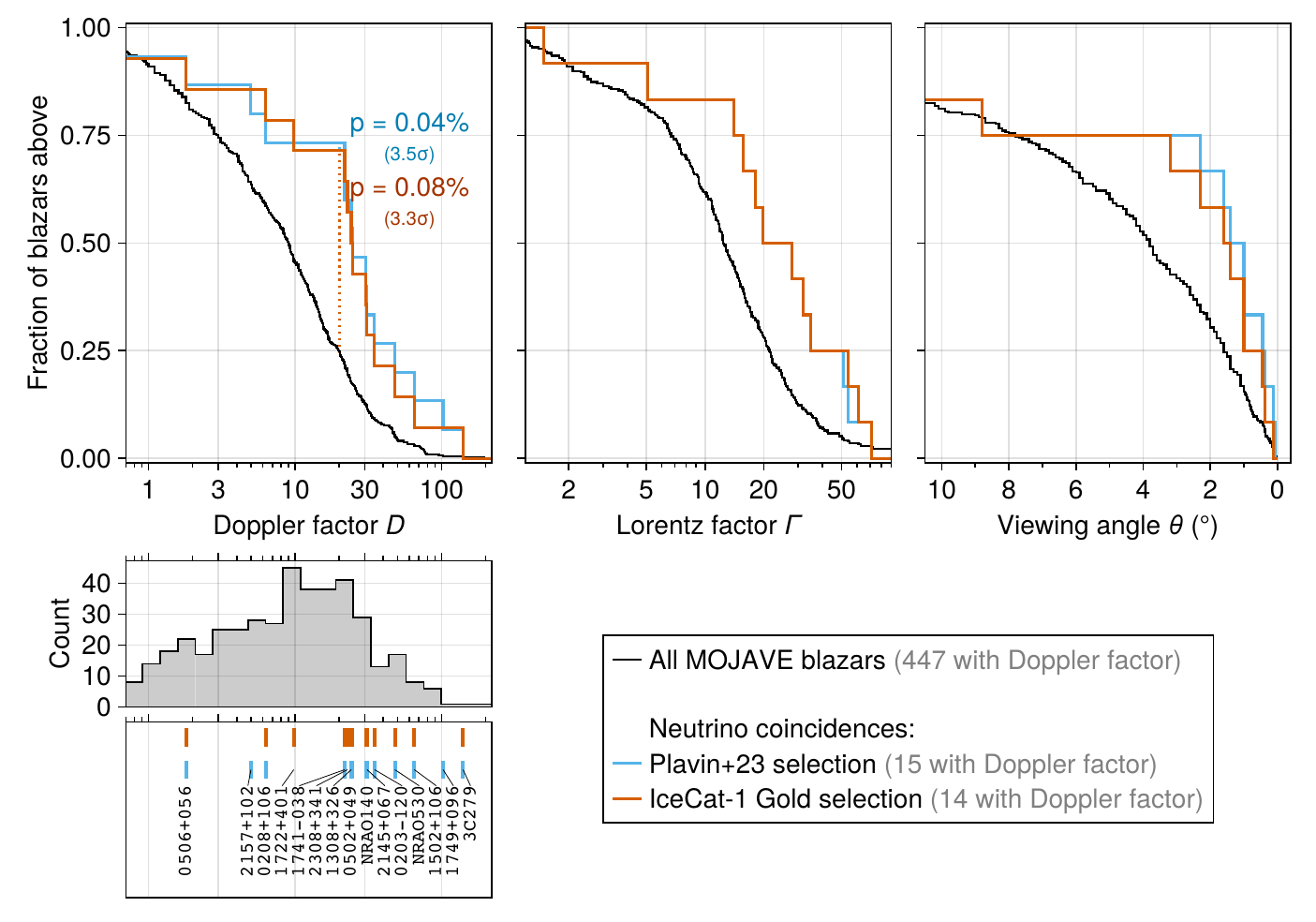}
\caption{Comparison of observational jet beaming indicators (Doppler $D$ and Lorentz $\Gamma$ 
factors, viewing angle $\theta$) for neutrino-coincident blazars and for the full sample of blazars observed in MOJAVE. For comparison, see \autoref{f:nudoppler} that only shows 1.5JyQC sources. \label{f:nudoppler_all}}
\end{figure*}

In the main analysis of this paper, \autoref{s:beaming-obs}, we utilize the uniformly-selected subsample of the full set of blazars observed within the MOJAVE program (1.5JyQC). That subset is purely flux-density limited (\autoref{s:data_vlbi}), leading to the most robust findings. The VLBI images of the nine 1.5JyQC blazars coincident with neutrinos are shown in \autoref{f:images}. Still, it is potentially instructive to look at the full MOJAVE blazar sample consisting of 447 sources with measured Doppler boosting \citep{MOJAVEdoppler}. The comparison is shown in \autoref{f:nudoppler_all}. The presence of relatively fainter blazars makes the neutrino-coincident ones stand out even more than in the 1.5JyQC analysis (\autoref{f:nudoppler}). This effect highlights the importance of population studies, large-sample survey and monitoring programs.

\end{document}